\documentclass[preprint,12pt,authoryear]{elsarticle}

\usepackage[utf8]{inputenc}
\usepackage[T2A]{fontenc}
\usepackage{subcaption}
\usepackage{doi}

\usepackage{hyperref}
\usepackage{xurl}

\usepackage{amssymb}
\usepackage{amsthm}
\usepackage{amsmath}
\usepackage{bm}
\usepackage{nicefrac}
\usepackage[textsize=small]{todonotes}

\journal{arXiv} 

\newcommand{\overbar}[1]{\overline{\!#1}}

\begin{document}

\begin{frontmatter}

\title{Ukrainian-style oligarchic economies:\\
how concentrated power undermines \\
value added in production chains
}

\author[inst1]{Jakub Karnowski}

\affiliation[inst1]{organization={SGH Warsaw School of Economics},            addressline={al. Niepodleglości 162},
            city={Warszawa},
            postcode={02-554},
            country={Poland}}

\author[inst1]{Przemysław Szufel}

\begin{abstract}

Oligarchic control exerts significant distortions on economic efficiency. Ukraine exemplifies this phenomenon, where oligarchs dominate key sectors and achieve economies of scale through vertical integration of coal mines, steel mills, and power plants while controlling critical infrastructure (e.g. access to transportation networks) to stifle competition. Their Soviet-era production chain monopolization strategies, coupled with political patronage networks (including both local and national governments), reinforce systemic inefficiencies and barriers to market entry.

Although existing studies highlight the developmental benefits of de-oligarchization, this work advances the literature through computational modeling. We develop an agent-based model of a partially oligarch-controlled economy, where firms with heterogeneous production functions interact within a value-added network. Through numerical simulations, we quantify how different de-oligarchization policies affect aggregate GDP growth.

The results indicate that the optimal de-oligarchization strategies are determined by the position of the oligarch in the production chain. Depending on the oligarch's position, dismantling oligarchic structures should either focus on removing oligarchs' access to raw materials or on breaking oligarchs' influence on other transactions in the production chain.

\end{abstract}

\begin{keyword}
oligarchs \sep de-oligarchization \sep oligarchic control \sep production chains \sep oligarchic economies \sep computational economics
\end{keyword}

\end{frontmatter}


\section{Introduction}
\label{sec:intro}

In this paper, we focus on \emph{oligarchs}  -- entrepreneurs who use their wealth to exert
political influence \citep{pleines2016oligarchs}, get more favorable regulations for their businesses \citep{winters2009oligarchy, pleines2016oligarchs}. Such regulations can heavily limit access to the other market participants \citep{stigler197regulation}. However, in post-Soviet countries, oligarchs have a more significant and deeper influence on market processes and the entire economy compared to Western countries.

For illustrative purposes, we focus on the Ukrainian-style economy. 
We define a \emph{Ukrainian-style economy} as an economic system in which a small group of powerful individuals or families (oligarchs) control a significant portion of the country's wealth, resources, and industries. In such a market, several economic sectors are entirely monopolized by oligarchs, while in other sectors they can still influence how transactions between market actors are concluded. This extensive control allows oligarchs to shape political and economic decisions to benefit their own interests rather than those of the broader population. In these economies, growth potential is often overshadowed by financial mismanagement, external dependencies, and systemic corruption. Since oligarchs, through their political influence, control regulatory decisions of local and state governments, the system exhibits persistent resistance to structural change. Note that the existence of such economies is not limited to post-Soviet contexts -- the approach presented in this paper can be applied to any economic system with these characteristics.

\citet{wilson2021faltering} points out that in Ukraine an anti-oligarch bill with a formal definition of an oligarch has been introduced by the administration of President Volodymyr Zelensky in 2021. Under this bill, an oligarch is a person who meets at least three of the following four criteria: (1) participates in public life; (2) has a significant impact on the media; (3) is the ultimate owner or controller of local monopolies; and (4) has confirmed financial assets worth more than
\$83 million. The author notices that the fourth criterion includes more than 100 individuals in Ukraine. \citet{lankina2019soviet} discuss how Ukrainian oligarchs, by owning industrial enterprises with huge employment, have direct control over a vulnerable and dependent workforce. This control means political power, as the oligarch is able to influence election outcomes. The authors introduce an electoral quality index and show that it is negatively correlated with the number of oligarchs in Ukrainian oblasts. The notable influence of oligarchs on Ukrainian politics is also confirmed in other research, e.g. \citet{kobzova2015ukraine, pleines2016oligarchs}.

The role of oligarchy in the Ukrainian economy is immense. \citet{goriunov2023oligarchic} point out that at the beginning of 2021, the oligarchs in Ukraine owned 36 of the 100 largest private enterprises. Further, they note  oligarchs owned  $12\%$ of total assets of Ukrainian companies with shares of revenues and employment $9\%$ and $4\%$, respectively. According to their report, industries with oligarchic ownership exceeding $50\%$ -- ranked from highest to lowest degree of oligarchization -- are: coke production, iron ore mining, coal mining, metal tubing, oil refining, electricity distribution, base metals, agricultural fertilizers, airlines, construction, television, gas distribution, and confectionery production. It should be noted that, in the case of heavy industries, the oligarchs control the entire production chain from raw materials to final products. In an earlier research \citet{wilson2016survival} noted that 50 richest Ukrainians own $45\%$ of the country's GDP while a similar number is $20\%$ in Russia and less than $10\%$ in the US. This huge development of the Ukrainian oligarchy is the result of the not transparent and badly executed privatization process that took place in the 1990s and the weakness of the Ukrainian state \citep{horiunov2022oligar}. \citet{vatamaniuk2022economy} points out that after the collapse of the Soviet bloc at the end of the 1980s, different speeds of transformation from a centrally planned economy to a market economy led to different strengths of the state. He notices that due to the weakness of the state combined with undeveloped civil societies led to \emph{state capture} and the formation of an oligarchic economy. \citet{pleines2016oligarchs} points out that this is even worsened by the fact of political control of oligarchs over the economy.

The process of reducing the role of oligarchy in the Ukrainian economy is difficult. \citet{wilson2016survival} noted that while corruption was widespread in Ukraine, the major blocker for any reforms is \textit{``the inter-penetration of the corrupt political class and the superrich oligarchy''}. However, in 2021 an anti-oligarchic law was passed. If someone is deemed an oligarch, they are barred from funding political parties and participating in the privatization of valuable state assets. They must also file declarations as public officials, the so-called e-declaration, which the coauthor of this paper filed as well as an independent supervisory board member of SOEs. The law also proposes adding oligarchs (and possibly their companies) to a high-risk list for financial monitoring. \citet{horiunov2022oligar} point out that once the law has been passed, the oligarchs are already identifying new regulatory loopholes to circumvent those regulations.

The goal of this paper is twofold: firstly to explore to what extent the negative effects of oligarchs can lead to a decrease in the overall GDP of an economy, and, secondly, to analyze the potential pathways for de-oligarchization. This is achieved using a stylized economic model that captures the effects of the network on the value-added chains of an economy.  It has been observed that oligarchs are particularly interested in controlling large sets of vertically integrated companies \citep{lankina2019soviet, gorodnichenko2008oligarchs}. Creating holdings gives them a set of tools that can be used to maximize their profits. In particular, via offshore vehicles, oligarchs can efficiently avoid taxes. Moreover, tax evasion can be taken as a coordinated action by a group of oligarchs and can lead to magnified economies of scale (compared to the tax evasion capabilities of individual enterprises). Furthermore, oligarchs can untransparently redistribute resources within their holdings to further increase profits and avoid taxes \citep{gorodnichenko2008oligarchs}.

In our model, we will show how integrated production chains controlled by oligarchs can affect the economic efficiency of the entire economy.
In order to capture this complexity, we will take \emph{system approach} \citep{ackoff1971towards}.  An economy can be represented as a system of interconnected entities whose actions yield added value that is subsequently shared using various mechanisms (taxes, monetary transfers, regulation, etc.) among the system's participants. In a market system, different actors on the market have different powers and different impact on the overall productivity that depends on their position in the added value chain. Hence, we need an approach that allows us to model heterogeneity of actors along with capturing network dependencies. The approaches that allow us to model such systems include complex system theory \citep{miller2009complex} and agent-based simulation \citep{macal2009agent, tesfatsion2002agent}. Typically, possible interactions between market participants are typically represented as networks \citep{tesfatsion2006handbook}. Similarly, we will use networks to represent production chains within the economy, with nodes representing manufacturers and edges representing flows of goods. This will allow us to observe the cascading effects of the behavior of oligarchs \citep{baqaee2018cascading}. Actions of a single oligarch can result in negative effects for several connected companies and can lead to inefficiencies on a scale far exceeding the profits of an oligarch. \citep{vitali2011network} discusses cascading failure in network-based systems.

In this paper by \emph{oligarchization} of an economy, we will understand the level at which the economic processes are being controlled by the oligarchs. \citet{vatamaniuk2022economy} propose to use as the oligarchization the proportion of oligarchs' wealth to the entire GDP of the economy. To be precise, in the text we will further interpret as oligarchization the percentage of companies in the modeled total value-added chain that are being owned and controlled by an oligarch. 

This paper extends the existing literature in several ways. The previous literature focuses on a detailed description of the current oligarchization state along with comparative analysis of economies that have taken different paths after the collapse of the communist regime. For example, \citep{akerman2016oligarchies} are considering a two-sector economy having agriculture and manufacturing production, represented by Cobb-Douglas functions, where oligarchs control access to capital and land. We will extend this approach by considering a network of companies connected in a production chain where parts of the chain are controlled by oligarchs. In another related paper, \citet{wright2015} show a model of a western-styled economy where oligarchs can influence the elections by funding the political parties and in the results get more favorable tax laws that allow them to avoid paying taxes. They show that individual actions of the system's participants (particularly voters) lead to a situation where, in fact, the oligarchs have significant influence on both the final election results and the changes in tax laws, policy making, and market regulations.   In this work, we take a wholly new approach by constructing a stylized model of the oligarchic network economy that focuses on value-added chains. Our approach is inspired by agent-based computational economics, where interactions between companies are modeled as a network \citep{tesfatsion2002agent}. 
However, to the best of our knowledge, no stylized network model of the Ukrainian-type oligarchic economy has been proposed in the literature. The contributions of our paper are as follows.
\begin{enumerate}
    \itemsep0.3em
	\item a new class of a stylized model of an oligarchic economy based on a network of companies connected in a production chain;
	\item analysis on how different oligarchization levels are affecting via the network effects the overall added value of an economy;
	\item understanding how the location of an oligarch in the production chain can influence the overall added value of an economy;
	\item demonstrating when de-oligarchization can have either positive or negative effects on the output of an economic system;
	\item recommendations for de-oligarchization policy.
\end{enumerate}

The paper is organized as follows. After this introduction, in Section~\ref{sec:literature} we discuss the background of the literature on the influence of oligarchs on the economy. Subsequently, in Section \ref{sec:model} we present a stylized model of an oligarchic economy. We first start with a model of a production chain economy, and next introduce an oligarch. Next, in Section \ref{sec:experiments}, we present and discuss the results of numerical experiments. Finally, in Section \ref{sec:conclusions}, we draw conclusions.

\section{Literature review}
\label{sec:literature}

\subsection{Oligarchs and oligarchy}
\label{sec:oligarchs_oligarchy}

The concept of oligarchy, traditionally understood as the rule of the few, has evolved significantly in political and economic discourse. \citet{aristotle1944politics} described oligarchy as governance controlled by the wealthy minority, in contrast to democracy, which he characterized as rule by the poor majority. \citet{plato380republic} similarly viewed the oligarchy as a system in which political power is concentrated in the hands of those who possess substantial property, thereby excluding the less affluent from decision-making. His critique emphasized that oligarchic structures arise due to inadequate education and governance, leading to systemic inequalities \citep{plato380republic}.

Robert Michels \citeyear{michels1915political}  in his ``Iron Law of Oligarchy'', points out that even organizations initially founded on democratic principles inevitably develop oligarchic tendencies as a small leadership group consolidates control over decision-making processes \citep{shaw2014laboratories}.
\citet{mills1956power} offers a complementary view, emphasizing the role of the \emph{``power elite``} -- a small group of individuals who, by virtue of their economic or institutional positions, exert disproportionate influence over societal outcomes. 
Albert Einstein \citeyear{einstein1949socialism} extends this discussion into the economic realm, arguing that capitalism inherently leads to oligarchic structures due to the natural tendency of private capital to become concentrated in fewer hands. He attributes this phenomenon to both competitive pressures and the efficiencies derived from large-scale industrial production, which systematically disadvantage smaller market participants. 

Moving to the current economic dispute, \citet{hartmann2021hidden} echoes this concern, describing how economic elites translate financial power into political influence, often by funding political campaigns or lobbying efforts to shape regulatory environments in their favor. This process gradually erodes democratic safeguards, facilitating the transition from democracy to oligarchy.

\citet{winters2011oligarchy} further refines the definition by distinguishing oligarchs from general elites, stating that while elites may derive power from political, social, or military influence, oligarchs exert control primarily through wealth accumulation and its strategic use to shape policy and governance. His argument challenges the assumption that democracy inherently displaces oligarchy, instead suggesting that democratic institutions often become intertwined with oligarchic interests, leading to a fusion rather than a displacement of power \citep{winters2011oligarchy}. \citet{tabachnick2011oligarchy} reinforce this perspective, noting that the classical notion of oligarchy has regained relevance in global politics, particularly in discussions of economic inequality and elite capture.

Oligarchic influence is exerted through various mechanisms, including direct control over political institutions, lobbying, and informal negotiations with policymakers. \citep{nastain2023failure} identify lobbying and political negotiations as primary tools through which oligarchs shape government policies, ensuring that regulatory frameworks serve their interests. Furthermore, oligarchs may not always seek direct political office; instead, they often operate behind the scenes by financing political parties or maintaining strategic alliances with policymakers \citep{nastain2023failure}.

This covert influence is particularly evident in post-industrial economies, where the regulatory landscape is complex and susceptible to elite manipulation. \citet{acemoglu2008oligarchic} warns that once oligarchs establish economic dominance, they work to institutionalize their advantages by restricting access to critical resources such as financial capital, infrastructure, and technology. This creates a self-reinforcing cycle in which oligarchs leverage their wealth to perpetuate their control, reducing overall economic dynamism and social mobility.

\citet{tabachnick2011oligarchy} suggest that global economic trends, particularly in the post-Cold War era, have exacerbated oligarchic tendencies. The not transparent privatization of state assets, deregulation of financial markets, and globalization of capital flows have collectively enabled a new class of transnational oligarchs to emerge. This contemporary oligarchy transcends national borders, operating through multinational corporations and offshore financial networks -- further complicating regulatory efforts.

\citet{aristotle1944politics} classical observation that the rich are always the few and the poor the many remains a fundamental principle in modern discussions of economic inequality. The persistence of oligarchic structures in different political systems underscores the challenges of achieving truly equitable governance. The theoretical and empirical literature reviewed here highlights the mechanisms through which oligarchs consolidate power, the economic consequences of their dominance, and the potential pathways through which societies can counterbalance these influences. 

Comparative historical analysis suggests that oligarchic dominance is neither absolute nor permanent. As \citet{acemoglu2008oligarchic} illustrates, while oligarchic societies can initially experience economic success, their resistance to competition and institutional innovation ultimately leads to stagnation. The decline of oligarchic regimes is often precipitated by internal inefficiencies, external economic pressures, or political movements that advocate for redistributive reforms.

\subsection{Oligarchic economy}
\label{sec:oligarchic_economy}

The oligarchic economy is an economic system in which a small group of individuals, families, or economic groups control a large portion of wealth and combined economic and political power.  Concentrated control enables them to influence the legislature, political decisions, economic policies, and market practices to safeguard and expand their influence. 

The relationship between oligarchy, economic inequality, and systemic instability has been a central topic of both classical and modern economic thought. Oligarchy negatively affects aspects of everyday life, such as politics, market, inequality, economy, social life, and democracy. The concentration of power in the hands of a small group of people makes it difficult to implement reforms to boost economy and favors the interests of oligarchs at the expense of the public good, leading to corruption and nepotism, as well as inequality, as a manifestation of social injustice, illegal activities of wealthy individuals or activities of hidden interest groups influences both politics and democracy. We discuss those areas in subsequent sections of this chapter.

\subsubsection{Political influence }
\label{sec:political_influence}

Oligarchs wield political influence through both incentives and coercion. As documents \citet{hartmann2021hidden}, they reward politicians and parties aligned with their interests while systematically undermining those who resist. This dual strategy -- financing compliant officials while marginalizing or crushing opposition -- ensures governance structures remain subservient to elite interests.

The ability of oligarchs to shape policy extends beyond direct political contributions; they also control the mechanisms of policy formulation through lobbying and negotiation \citep{nastain2023failure}. By embedding themselves within government decision-making processes, oligarchs ensure that policies are tailored to their benefit, rather than serving the broader public interest.

As oligarchs consolidate control over political and legal institutions, the rule of law weakens, making it increasingly difficult to enforce property rights and maintain a fair economic environment. \citet{bourguignon2000oligarchy} argue that in oligarchic systems, strong protection of property rights is unlikely to emerge, as elites benefit from maintaining an environment of legal uncertainty. This leads to increased economic risk, discouraging long-term investments and innovation.

One of the most insidious forms of oligarchic influence is its control over information. \citet{hasan2024critical} argue that when a small number of corporations control the media industry, they inevitably dominate public discourse. This control allows oligarchs to suppress dissenting narratives, manipulate public opinion, and reinforce ideological frameworks that legitimize their dominance. In Ukraine, for example, oligarch-owned media outlets have historically shielded their benefactors from scrutiny while attacking political opponents \citep{tarasyuk2021stratagems}.

\subsubsection{Market entry barriers}
\label{sec:market_entry_barriers}

Market entry barriers are obstacles that make it difficult for new competitors to enter an industry or market. Examples are economies of scale, high capital requirements, access to distribution channels, various government regulations such as licensing and permits, and patents. 

\citep{stigler1983organization} defined market entry barriers as \emph{``a cost of producing that must be borne by a firm which seeks to enter an industry but is not borne by firms already in the industry''}. Another American economist \citep{bain1956barriers} gave the definition of entry barriers as \emph{``an advantage of established sellers in an industry over potential entrant sellers, which is reflected in the extent to which established sellers can persistently raise their prices above competitive levels without attracting new entrants to enter the industry''}.

In contemporary political economy, \citet{acemoglu2008oligarchic} highlights that oligarchic societies initially experience economic growth due to the protection of property rights for dominant producers. However, these societies tend to stagnate over time as entrenched elites erect barriers against new market entrants, thereby stifling innovation and broader economic participation. The cyclic nature of the rise and decline of oligarchs suggests that, while oligarchic control can drive short-term economic expansion, it ultimately results in inefficiencies and comparative decline when juxtaposed with democratic governance models \citep{acemoglu2008oligarchic}.

The overwhelming majority of public utility providers, such as electricity and water services, are privatized and controlled by a small circle of individuals. Thus, every increase in utility tariffs redistributes income from small private businesses that produce essential goods to the pockets of oligarchs who own natural monopolies in the country. This leads to the destruction of the market economy and the formation of an economic system of state monopolism.
In addition, oligarch monopolization of key sectors of the economy is one of the reasons for the poor investment climate. The dependence of most political forces on business means that in many cases state authorities are guided not by the interests of the country, but by the oligarchs who sponsor them. 
\citep{snyder2021tyranny} mentions that \emph{``we certainly face, as did the ancient Greeks, the problem of oligarchy -- ever more threatening as globalization increases differences in wealth''}. Given the high degree of concentration of ownership in the hands of relatively few oligarchs, there is a high probability that a number of strategic enterprises may go bankrupt.

\citep{matuszak2012democracy} points out that the main weaknesses of a monopoly or oligopoly are limited human resources, which the oligarch may consider unconditionally loyal to him, and the strength of the remaining oligarchic groups. Unlike perfect competition, where many small businesses operate in the market, in an oligopoly, a few large entities control the market and have a significant influence on prices and the quantity of products offered.

\subsubsection{Inequality}
\label{sec:inequality}

The accumulation of extreme wealth among oligarchs results in a disproportionate concentration of political and economic power, exacerbating social disparities, and distorting market mechanisms. \citet{bilan2020impact} provide empirical evidence that shows that extreme income disparities correlate with weaker economic and social development. Countries with a more balanced income distribution, reflected in lower Gini coefficients, tend to achieve higher levels of human development and economic resilience. As \citet{winters2011oligarchy} asserts, extreme material inequality directly produces extreme political inequality, reinforcing the structural imbalances that privilege oligarchic elites at the expense of greater economic participation. This economic stratification is not a passive phenomenon, but a dynamic process in which those who control substantial wealth actively manipulate institutional frameworks to perpetuate their dominance. \citet{nastain2023failure} emphasize that economic inequality itself serves as a defining marker of oligarchic rule, where material resources function not just as economic assets, but as political commodities that sustain elite control.

Empirical studies further demonstrate that the degree of oligarchization within an economy is significantly correlated with patterns of wealth distribution and social mobility. Michels’ (\citeyear{michels1915political}) thesis on oligarchic entrenchment within political organizations finds parallels in economic structures, where a small group of dominant firms or financial actors exercises control over key markets. \citet{mills1956power} similarly notes that power consolidation in business and politics reinforces social stratification, limiting opportunities for upward mobility among lower-income groups.

Persistent inequality has profound implications for economic efficiency and long-term growth. \citet{Stiglitz2011ofthe1perc} underscores that high inequality weakens overall economic productivity by reducing aggregate demand. When wealth is concentrated in the hands of a few, consumption declines as the spending capacity of lower-income groups diminishes, leading to underutilization of productive resources. This demand deficiency generates instability, further exacerbating inequality and creating a self-reinforcing cycle of economic stagnation. A similar argument is made by \citet{sonin2003rich}, who describes how economies with entrenched oligarchic control often fall into a low-growth equilibrium characterized by income disparities and widespread rent-seeking behaviors. These structural inefficiencies hinder innovation and limit the potential for inclusive economic expansion.

Furthermore, inequality distorts market competition by privileging incumbents and suppressing new entrants. \citet{vries2017decline} points out that when factors of production become commodified, those who already control a disproportionate share of assets gain a compounding advantage, leading to further concentration of wealth. This dynamic restricts market competition, misallocates resources, and suppresses entrepreneurial activity, ultimately suppressing long-term economic growth. Similarly, \citet{stiglitz2012price} explains that monopoly power exacerbates inequality by generating excessive rents, which in turn produce economic distortions. Consumers face higher prices, leading to inefficient allocation of resources as demand shifts away from monopolized goods and services. In such environments, investment decisions are not dictated by competitive market forces, but rather by entrenched oligarchic interests seeking to preserve their dominance.

\subsubsection{Economic growth}
\label{sec:economic_growth}

Oligarchic dominance extends beyond the realm of economics, shaping political institutions, social values, and public discourse in ways that undermine long-term national development. The concentration of wealth and power in the hands of a small elite not only distorts markets but also erodes democratic institutions and weakens civil society. Over time, this process contributes to declining economic dynamism, weakened property rights, and systemic corruption, ultimately leading to social stagnation and institutional decay.

One of the most damaging economic consequences of oligarchy is its tendency to inhibit competition and stifle market development. \citet{robinson2012nations} argue that political elites with unchecked economic power will prioritize rent-seeking over economic progress, using their influence to block competition and secure monopolistic privileges. These outcomes lead to inefficient organizational frameworks in business, where private ownership is dictated not by market effectiveness but by the ability to utilize appropriated resources.

Oligarchic dominance not only perpetuates inequality, but also actively distorts key economic indicators. \citet{bourguignon2000oligarchy} argue that highly unequal societies experience slower democratization and institutional development, reinforcing systemic inefficiencies. In oligarchic economies, market transactions are often manipulated to benefit entrenched elites. \citet{gorodnichenko2008oligarchs} note that oligarchs frequently overstate costs and understate sales, artificially reducing measured productivity. This manipulation further entrenches inequality by shifting wealth from productive economic actors to those who exert monopolistic control over key industries.

\subsubsection{Social erosion}
\label{sec:social_erosion}

Beyond its economic and political effects, the oligarchy also has profound social consequences. \citet{tarasyuk2021stratagems} note that the rule of oligarchy erodes the moral and ethical foundations of society, fostering a culture in which success is measured by proximity to power rather than merit or innovation. This shift in value orientations discourages civic engagement and reinforces public cynicism toward democratic institutions.

The financialization of the economy under the oligarchic rule further accelerates social fragmentation. \citet{vries2017decline} warns that as wealth concentration intensifies, it erodes the organizational norms and institutions that once supported broader economic participation. Worse still, once this process becomes entrenched, it is nearly impossible to reverse through conventional political means, as economic elites capture decision-making structures.

\citet{dabrowski2017ukraine} mentions that the oligarchic system creates deep social distrust towards the government and undermines the legitimacy of the entire political system. Even when politicians make the right decisions, they are suspected of doing so for their own material benefit. Studies also highlight that one of the consequences of state capture is a decline in public trust in state institutions, which leads to a further deepening of the socio-economic crisis in the country.

\subsubsection{Erosion of democracy}
\label{sec:societal_decline}

The rule-of-law problem is central to understanding oligarchic persistence. \citet{winters2011oligarchy} argues that in many societies, governance challenges are not about eliminating oligarchy but rather about mitigating its excesses through legal frameworks. This is particularly relevant in hybrid political systems, where democratic institutions exist alongside deeply entrenched oligarchic networks. As a result, policymaking often reflects the interests of economic elites rather than the broader population.

\citet{cohen2024cycles} underscores that once oligarchic systems take hold, they become highly resistant to reform. Existing democratic institutions are often deliberately designed to prevent radical disruptions to the status quo, making it difficult to dismantle oligarchic power without extreme political upheaval. This creates a paradox where democratic mechanisms coexist with oligarchic control, rendering traditional checks and balances ineffective.

\citet{hartmann2021hidden} highlights how oligarchs use their financial power to manipulate political institutions, funding media outlets, lobbyists, and think tanks to dominate public discourse. By embedding themselves in popular culture and academia, they systematically weaken regulatory safeguards designed to keep wealth out of politics. Over time, this influence erodes democratic institutions, making it nearly impossible to reverse oligarchic dominance through conventional political means \citep{vries2017decline}.

The control of public institutions by oligarchs also weakens democratic governance, as elections become a battleground for elite interests rather than genuine representation of the populace. \citet{nastain2023failure} note that democratic systems that rely on direct elections are particularly vulnerable to corruption, as oligarchs use their wealth to buy influence and manipulate electoral outcomes. This process transforms nominal democracies into \emph{``embryonic oligarchies''} where political competition exists in form but not in substance.

\subsection{The case of Ukraine}
\label{sec:ukraine}

The oligarchization of the Ukrainian economy has had a profound impact on economic growth since independence in 1991. Before Euromaidan in 2014 as well as before the full-scale Russian invasion in 2022, oligarchs controlled key sectors of the economy: energy, banking and media, distorting competition. Although they have brought some investment and capital, their influence has slowed reforms, created inefficiencies, and fueled corruption by using political connections to secure favorable conditions, ultimately slowing economic growth and diminishing the state capacity to defend itself against the Russian invasion. Therefore, it is of crucial importance for country policymakers to fully understand how oligarchization of the Ukrainian economy influences Ukraine's long-term growth and the prosperity of the Ukrainian people. 

\subsubsection{Oligarchization of the Ukrainian economy}
\label{sec:oligarchization_ukraine}

The oligarchization of the Ukrainian economy can be traced back to the country’s rapid and poorly regulated privatization process following the collapse of the Soviet Union. This transition created an environment highly conducive to rent-seeking behavior, where a select group of individuals acquired vast economic assets, often for the purpose of asset stripping rather than productive enterprise. These oligarchs consolidated their power by exerting significant influence over political institutions, regulatory frameworks, and key sectors of the economy. The result was an extensive system of monopolization that persisted for decades, distorting market mechanisms and reinforcing economic inequalities \citep{bilan2020impact}.

The dominance of Soviet-era firms further exacerbated inefficiencies in the early years of the Ukraine transition to a market economy. \citet{gorodnichenko2008oligarchs} note that the large enterprises inherited from the Soviet system were notably unproductive, failing to adapt to the competitive pressures of a free market. Instead of fostering innovation and efficiency, these firms often relied on oligarchic networks for preferential treatment, subsidies, and regulatory protections. \citet{simonchuk2024social} argues that this symbiotic relationship between political power and economic dominance entrenched oligarchs as the foundational social class within Ukraine, effectively marginalizing small and medium-sized enterprises (SMEs).

A defining feature of Ukraine’s oligarchic system was the role of the media in shaping public perception and shielding oligarchs from accountability. \citet{tarasyuk2021stratagems} highlight how oligarch-controlled media outlets systematically manipulated public opinion, obscuring corrupt dealings, and limiting political opposition. \citet{kuznetsov2023ukraine} extends this analysis, highlighting that the oligarchic class not only controlled vast portions of Ukraine's wealth but also key communication channels, reinforcing their dominance through information asymmetry. As a result, economic reforms faced considerable resistance, as political and legal institutions remained deeply intertwined with elite interests.

\subsubsection{De-oligarchization of the Ukrainian economy}
\label{sec:deoligarchization_ukraine}

The de-oligarchization of Ukraine represents one of the most significant economic and political transitions in post-Soviet Europe. Although oligarchs have historically dominated Ukraine’s economy and political landscape, recent developments, particularly the ongoing war and external economic pressures, have initiated a fundamental change in the economic structure of the country. However, this process is deeply intertwined with Ukraine’s broader struggles with corruption, monopolization, and economic inefficiencies. Scholars highlight the multifaceted nature of de-oligarchization, which involves both structural economic changes and political realignments that challenge entrenched elite dominance.

The Russian invasion of Ukraine has accelerated the de-oligarchization process, fundamentally reshaping the economic and political landscape of the country. \citet{rojansky2022} argues that the war has directly undermined the economic foundations of Ukraine’s oligarchic elite by disrupting their control over key industries, particularly in the occupied east of the country, where Russian aggression started. The loss of natural resources and industrial assets, historically monopolized by oligarchs, has forced the Ukrainian economy to pivot to new sectors, such as agriculture and information technology, where oligarchic influence is significantly weaker \citep{siedin2024}.

A particularly important consequence of this transformation is the diminishing role of oligarchs in strategic economic sectors. \citet{siedin2024} notes that multimillionaires in emerging industries no longer control monopolies over critical resources, such as gas and electricity, nor do they dominate financial institutions. This structural shift limits their ability to convert economic power into political influence, thereby reducing their ability to manipulate regulatory frameworks in their favor. The weakening of the oligarchic class is thus not merely a political development but a fundamental restructuring of Ukraine’s economic hierarchy.

This shift is also evident in the declining influence of oligarchic media networks. As war-related disruptions have undermined their financial stability, many oligarchs have been forced to withdraw from their media holdings, thus reducing their capacity to shape public discourse \citep{rojansky2022}. 

With declining economic leverage and weak political connections, the traditional oligarchic model is increasingly untenable in the evolving economic order of Ukraine.

Although war-induced disruptions have significantly weakened Ukraine’s oligarchic system, long-term de-oligarchization requires sustained institutional reforms. \citet{kuznetsov2023ukraine} highlights that the elimination of corruption and the strengthening of the rule of law are essential preconditions for the success of international aid efforts, which have become a crucial component of Ukraine’s economic resilience. Without such reforms, there is a risk that old oligarchic structures could re-emerge in new forms, perpetuating existing inefficiencies.

\citet{rojansky2022} further emphasizes that eliminating oligarchic influence is critical for aligning Ukraine with European economic and governance standards. He argues that the dismantling of monopolistic control and the promotion of transparent regulatory frameworks will facilitate the integration of Ukraine into European markets, making it easier to adopt international best practices. \citet{simonchuk2024social} reinforces this perspective, asserting that only by successfully integrating into European institutions can Ukraine fully transition away from an oligarchic economic model. This process would involve economic liberalization, increased protections for SMEs, and a shift toward a more competitive and transparent business environment.

However, a key challenge lies in overcoming deeply ingrained social and economic behaviors that have historically supported oligarchic structures. \citet{simonchuk2024social} points out that Ukraine's rent-oriented economic culture, where success is often determined by access to state resources rather than market efficiency, must be fundamentally reshaped. Without addressing this cultural dynamic, reforms may face resistance or be co-opted by new economic elites.

Despite recent progress, the complete dismantling of Ukraine’s oligarchic system remains an ongoing challenge. The persistence of monopolistic tendencies and rent-seeking behaviors poses a significant risk to Ukraine’s long-term economic development.

However, the changing economic structure, driven by war-related disruptions and external financial support, provides a unique opportunity for transformation. \citet{siedin2024} argues that as new economic sectors emerge and gain prominence, they will create alternative power centers that dilute the influence of traditional oligarchs. If effectively leveraged, this transition could pave the way for a more diversified and resilient economic model.

Ultimately, Ukraine’s path to de-oligarchization is contingent on sustained political will, international cooperation, and a commitment to institutional integrity. \citet{rojansky2022} stresses that the removal of oligarchs from both the economy and politics is essential to establish a transparent and competitive economic system. This process will not only enhance Ukraine’s ability to integrate into European structures, but also foster long-term economic stability and social equity.

\subsection{Feedback loop of oligarchic inefficiency}
\label{sec:feedback_loop}

A feedback loop mechanism can be described as a process in which the outcome of an action influences its cause, which can lead to subsequent changes being strengthened or weakened. An example of a positive feedback loop is when oligarchs become wealthier and use their resources to gain even more power. A negative feedback loop, on the other hand, stabilizes a system, such as when growing social discontent reduces the influence of oligarchs through protests and reforms. In economics, a feedback loop occurs when the financial success of one group leads to even greater economic benefits for the same group.

Economic inequality does not operate in isolation; it has profound socio-political consequences that further entrench oligarchic dominance. \citet{Stiglitz2011ofthe1perc} contends that the increase in inequality is not just an economic phenomenon but also a reflection of the declining equality of opportunity. When economic mobility is restricted, entire segments of society are systematically excluded from contributing to and benefiting from economic growth. This inefficiency represents a fundamental misallocation of human capital, as significant portions of the population are unable to realize their productive potential.

Addressing the destabilizing effects of inequality requires both structural reforms and regulatory interventions. \citet{sonin2003rich} warns that without proactive measures, economies can become trapped in a low-growth equilibrium where inequality and stagnation reinforce each other. To break this cycle, policies must be designed to dismantle monopolistic structures and promote fair competition. \citet{stiglitz2012price}  suggests that progressive taxation, enforcement of anti-trust regulations, and targeted social investments are critical tools for mitigating inequality and restoring economic stability.

Moreover, inequality exacerbates political instability by eroding social cohesion. \citet{winters2011oligarchy} highlights that extreme wealth disparities generate extreme political imbalances, with policy decisions increasingly shaped by elite interests rather than democratic consensus. This imbalance manifests itself in slower institutional development and weaker governance structures, as highlighted by \citet{bourguignon2000oligarchy}, who argue that oligarchic societies democratize at a slower pace than more equal societies. \citet{bilan2020impact} further demonstrate that a more equitable income distribution correlates with higher indicators of social development, suggesting that addressing inequality is not only an economic imperative, but also a fundamental requirement for stable governance.

Ultimately, oligarchy fosters a society in which public institutions are repurposed to serve elite interests rather than the common good. This dynamic results in widespread disillusionment with governance, weakening social cohesion, and making collective action increasingly difficult. In such an environment, the prospects for economic and political renewal become severely constrained, locking societies into cycles of stagnation and inequality.

The persistence of economic inequality reinforces a vicious cycle in which economic and political power is increasingly concentrated. As \citet{gorodnichenko2008oligarchs} note, oligarchs often function as parasites, extracting wealth without generating additional economic value. This system of wealth redistribution towards elites leads to stagnation, as productive investments are replaced by rent-seeking activities. The resulting economic inefficiencies, coupled with declining institutional integrity, make it increasingly difficult for societies to implement effective redistributive policies.

In addition, institutional strengthening is necessary to counteract the political influence of entrenched elites. As \citet{bilan2020impact} demonstrate, equitable economic outcomes are strongly correlated with indicators of human development, suggesting that fostering inclusive growth is essential for sustainable development. Policymakers must also address distortions in financial and investment markets, ensuring that capital allocation prioritizes productivity and innovation rather than monopolistic rent-seeking.

\subsection{Conclusions}

The oligarchic economy system represents a complex interplay of economic, political, and social factors that perpetuate inequality and inefficiency. The oligarchs use their power and political influence to create barriers to entry for new market participants and stifle competition. The concentration of wealth and power in the hands of a few leads to inequalities that not only distort market mechanisms but also undermine democratic institutions and social cohesion. This complex dynamics fosters an environment where short-term rent-seeking by a few individuals prevails over productive investment, economic efficiency, and the welfare of the entire society. 
The negative outcomes of economy oligarchization are further increased due to the the feedback loop mechanism were the accumulation of wealth leads to greater political influence and stronger control of oligarchs over the economy. 

Hence, the oligarchs can influence the economy within the entire production chain, from the extraction of raw materials to the manufacturing of final goods, thereby shaping market dynamics and policy outcomes in their favor. On the other hand the market regulator is interested in increasing the economic efficiency and the welfare of the entire society. Since oligarchs typically control not single company but entire production chains, the regulator needs to understand how oligarchs influence the efficiency of the production chain by controlling raw resources or limiting the access of other companies to the market. This would allow to identify optimal regulatory interventions that can mitigate the negative effects of oligarchization. Such network model will be presented in the next section.

\section{Model}
\label{sec:model}

In this section, we describe the model of an oligarchic economy. We will take the approach to present the model similar to \citep{akerman2016oligarchies}: Firstly, we start with the description of the basic non-oligarchic model assumptions. Secondly, we introduce the oligarch in the model and extend the description to be able to understand the influence of oligarchs on the efficiency of a production chain.

\subsection{Baseline model assumptions}

Consider an economy comprising a set $K$ of goods, where each good is indexed by $k \in K$ with $k = 1,\dots,|K|$. Denote by $N$ the set of goods that are raw or natural resources that can be harvested for a given price. We will use the first $|N|$ indices of $k$, to denote the raw resources. That is $N = \{1, \ldots, |N|\}$ and $N \subset K$. Let us by $M$ denote a set of $|K| - |N|$ goods that are manufactured by the companies, that is, $M = \{ |N| + 1, \ldots, |K| \}$, obviously $N \cap M = \emptyset$ and $N \cup M = K$.

Each company produces exactly one good $m$, $m \in M$. In the text, we will use $m \in M$ to denote both goods and companies depending on the context. The production output $y_m$ is defined by a Cobb-Douglas  function:
\begin{equation}
    y_m = \alpha_m \prod_{k \in K}{x_{km}^{\beta_{km}}},
\label{eq:production_function}
\end{equation}
where:
\begin{itemize}
    \itemsep0.2em
    \item $y_m$ is the production output of the company $m$,
    \item $\alpha_m$ is the technology level of the company $m$,
    \item $x_{km}$ is the amount of good $k$ used for the production by the company~$m$,
    \item $\beta_{km}$ is the production coefficient of the good $k$ in the company $m$.
\end{itemize}
Note that when for some $k \in K$ and $m \in M$ the value of $\beta_{km}$ is zero, this means that the good $k$ is not used for the production of the company $m$. This means that sometimes in the production function we might have a term $0^0$. Throughout the paper, we assume $0^0 = 1$. Additionally, we assume that each company $m \in M$ has decreasing return to scale -- that is the production coefficients are such that:
\begin{equation}
  \sum_{k \in K}{\beta_{km}} < 1 \qquad \forall m \in M.
  \label{eq:decreasing_return}
\end{equation}
This can be caused by factors such as scarcity of resources or limited access to the market due to geographic location.

In addition, by $v_k$ we will denote the value of good $k \in K$ in the economy. For any given company $m \in M$, its value added $\rho_m$ can be calculated as a difference between the value of the output and the value of the inputs used to generate that output. That is:
\begin{equation}
      \rho_m = y_mv_m - \sum_{k \in K}v_kx_{km}.
\label{eq:value_added}
\end{equation}

Let us define the matrix $\bm{\beta}$ of production coefficients $\beta_{k_1k_2}$ as:
\begin{equation}
     \bm{\beta} = [\beta_{k_1k_2}]_{|K|\times|K|}.
\end{equation}
 For this definition, we assume $\beta_{k_1k_2} = 0$ for $k_1 \in K$, $k_2 \in N$. That is, in our model the natural resources are not produced but harvested or imported at a fixed cost $v_n$ where $n \in N$. Hence, the matrix $\bm{\beta}$ is quadratic and contains zeros in the first $|N|$ columns. Analogically, we define the matrix of good flows $\bm{x} = [x_{k_1k_2}]_{|K|\times|K|}$. Finally, we define a vector of values of goods $\bm{v} = [v_k]_{|K|}$.

Let us now define a weighted directed graph $G(K, E)$ where $K$ is the set of nodes representing goods in the economy and $E$ is the set of edges representing connections in the production chain. The adjacency matrix of the graph $G$ is given by $\bm{\beta}$. That is, the edge $(k_1, k_2) \in E$ if and only if $\beta_{k_1k_2} > 0$.
Further, we assume that $\bm{\beta}$ is defined in such a way that $G$ is a directed acyclic graph (DAG). Without loss of generality, we assume that the vertex indices of the graph $G$ are topologically sorted \citep{kahn1962topological}, that is, the matrix $\bm{\beta}$ is upper triangular. Finally, by $d(G)$ we will denote the depth of the directed graph $G$ as the maximum path length within the graph $G$.

Note that the nodes on the graph $G$ representing the raw materials, that is, $k \in N$ do not have any incoming edges. The nodes that do not have outgoing edges (that is, the nodes that have the corresponding row in the matrix $\bm{\beta}$ filled with zeros) represent companies manufacturing goods that do not participate in other production processes; we will code such goods \emph{final products}. The non-raw material goods that are used in other production processes (i.e., have at least outgoing edge) will be called \emph{intermediate products}.
In the model, we assume that both the intermediate and final products can be sold on the market at the price $v_k$. For $n \in N$ the value of $v_n$ represents the cost of acquiring raw materials.

\begin{figure}[ht!]
    \begin{minipage}{0.55\textwidth}
        \centering
        \includegraphics[width=\textwidth]{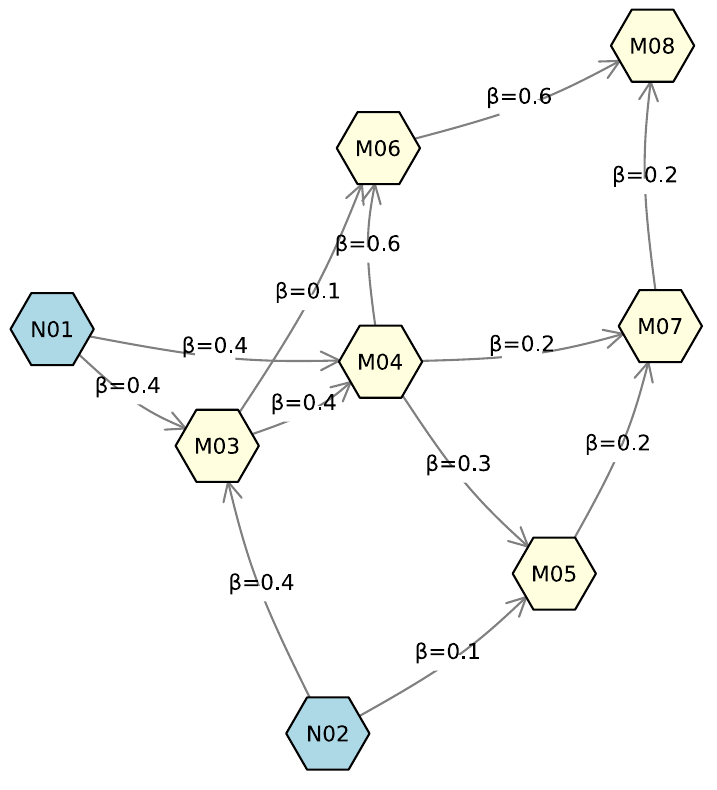}
    \end{minipage}
    \begin{minipage}{0.40\textwidth}
        \setlength{\arraycolsep}{3pt}
        \begin{equation*}
            \bm{\beta}  = \left[
            \begin{array}{cccccccc}
            0 & 0 & .4 & .4 & 0 & 0 & 0 & 0 \\
            0 & 0 & .4 & 0 & .1 & 0 & 0 & 0 \\
            0 & 0 & 0 & .4 & 0 & .1 & 0 & 0 \\
            0 & 0 & 0 & 0 & .3 & .6 & .2 & 0 \\
            0 & 0 & 0 & 0 & 0 & 0 & .2 & 0 \\
            0 & 0 & 0 & 0 & 0 & 0 & 0 & .6 \\
            0 & 0 & 0 & 0 & 0 & 0 & 0 & .2 \\
            0 & 0 & 0 & 0 & 0 & 0 & 0 & 0 \\
            \end{array}
            \right]
            \end{equation*}
    \end{minipage}
    \caption{A sample graph representing a production chain within an economy and the corresponding adjacency matrix $\bm{\beta}$. The nodes $N1$ and $N2$ represent natural resources, the nodes $M3,\ldots,M8$ represent companies manufacturing goods. The edges represent the corresponding $\beta_{km}$ parameters.}
    \label{fig:production_chain}
\end{figure}

Figure \ref{fig:production_chain} explains the approach taken in this paper by showing a sample graph representing a production chain within an economy and the corresponding adjacency matrix $\bm{\beta}$.  Note that the product produced by the company $M8$ is final - it does not participate in any other production processes, while the companies $M3,\ldots,M7$ manufacture intermediate products.

The total value added $\psi$ of the economy can be calculated as the sum of the value added of all companies in the economy. That is,

\begin{equation}
    \psi = \sum_{m \in M}{\rho_m} = \sum_{m \in M}{v_m y_m - \sum_{k \in K}v_k x_{km}}.
\end{equation}

In the model, we are considering no taxes other than transfers that might be needed to enforce the optimal production plan. Hence, if we consider that the described production chain constitutes the entire economy, the total value-added $\psi$ is equal to the GDP of the economy. 

The flows of goods in the economy $x_{km}$ that maximize GDP, that is, the total value added $\psi$ can be calculated as the solution of the following optimization problem:

\begin{equation}
\psi = \sum_{m \in M}\left(v_m \alpha_m \prod_{k \in K}{x_{km}^{\beta_{km}}}\right) - \sum_{k \in K}{v_k x_{km}} \rightarrow \max
\label{eq:goal}
\end{equation}
subject to the intermediate material availability constraints:
\begin{equation}
\sum_{m_2 \in M}{x_{m_1m_2}} - \alpha_{m_1} \prod_{k \in K}{x_{k{m_1}}^{\beta_{k{m_1}}}} \le 0 \qquad \forall m_1 \in M
\label{eq:balance}
\end{equation}

Denote by $\psi^*$ the value of the objective function \eqref{eq:goal} in the optimal solution and the flow matrix in that solution as $\textbf{x}^*$. Since the matrix $\bm{\beta}$ is upper triangular, the optimal flow matrix $\textbf{x}^*$ is also an upper triangular matrix. Finally, by $y^*_m$, $m \in M$ we will denote the amount of each good manufactured within the optimal solution.

In our model, we assume that the prices of raw materials, intermediate and final goods $v_k$, $k \in K$ are fixed and known by the market actors.
The optimal flow of goods $\textbf{x}^*$ maximizes that global added value $\psi^*$, but does not necessarily maximize the value added of each company defined in Equation~(\ref{eq:value_added}).
However, in our model, a production plan in which each company maximizes its value added is not Pareto-efficient for the entire economy. Hence, we assume that there is a central planner (market regulator) that is able to enforce the optimal production volume $y_m$ for each company in such a way that the global optimum $\psi^*$ is achieved. In real world this could be achieved by tax policy, subsidies, or other regulatory instruments.

\begin{figure}[ht!]
    \begin{minipage}{0.40\textwidth}
        \centering
        \includegraphics[width=\textwidth]{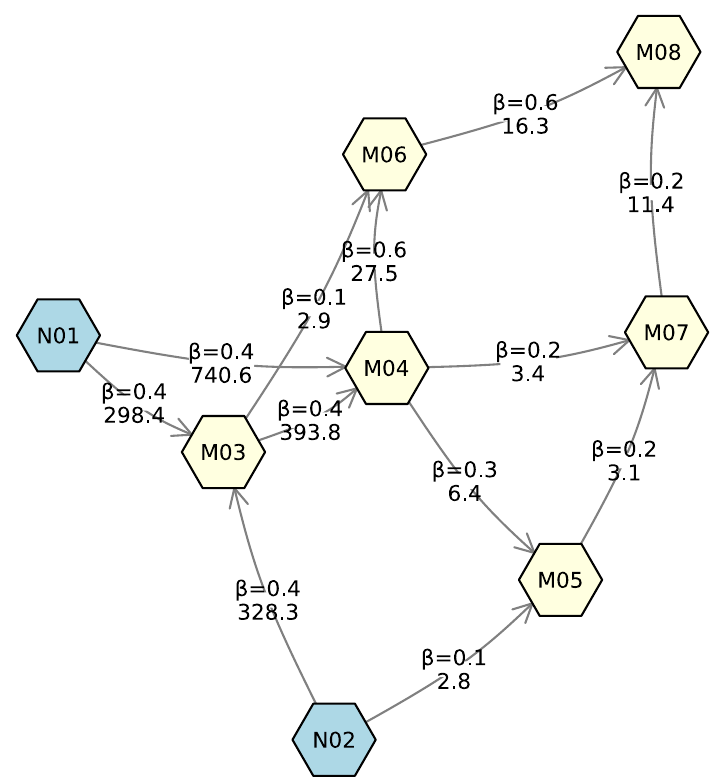}
    \end{minipage}
    \begin{minipage}{0.55\textwidth}
        \setlength{\arraycolsep}{3pt}
        \begin{equation*}
            \bm{x}  = \left[
            \begin{array}{cccccccc}
                0 & 0 & 298.4 & 740.6 & 0 & 0 & 0 & 0 \\
                0 & 0 & 328.3 & 0 & 2.8 & 0 & 0 & 0 \\
                0 & 0 & 0 & 393.8 & 0 & 2.9 & 0 & 0 \\
                0 & 0 & 0 & 0 & 6.4 & 27.5 & 3.4 & 0 \\
                0 & 0 & 0 & 0 & 0 & 0 & 3.1 & 0 \\
                0 & 0 & 0 & 0 & 0 & 0 & 0 & 16.3 \\
                0 & 0 & 0 & 0 & 0 & 0 & 0 & 11.4 \\
                0 & 0 & 0 & 0 & 0 & 0 & 0 & 0 \\
            \end{array}
            \right]
            \end{equation*}
    \end{minipage}
    \caption{An optimal production plan maximizing the $\psi$ as defined in Equation (\ref{eq:goal}) for the set of values $\bm{v} = [1.21, 1.1, 1.33, 1.46, 1.61, 1.77, 1.95, 2.14]^T$. Note that the flows are only present for non-zero $\bm{\beta}$ values.}
    \label{fig:production_chain_optimal}
\end{figure}

Figure \ref{fig:production_chain_optimal} shows a sample optimal production plan maximizing the $\psi$ found by solving the model defined in Equations (\ref{eq:goal}) and (\ref{eq:balance}). The technical details on how the solution was calculated are provided in Section \ref{sec:experiments}.

\subsection{Oligarch in the model}

We have defined a networked economy model with several manufacturers working jointly in a production chain towards a common added value.
In this section, we will extend this model by introducing an oligarch in the economy.

\citet{gorodnichenko2008oligarchs} point out that oligarchs tend to pick large companies for their holdings. They tend to create vertically integrated production chains. This is also confirmed in later research by \citet{pleines2016oligarchs} who also note that oligarchic companies have chosen a way to achieve vertical integration in various industries in order to maximize profit. \citet{gorodnichenko2008oligarchs} further note that oligarchs can own relatively non-profitable companies. This is also connected with the fact that profits of oligarchic companies are much more likely to be transferred to offshore vehicles. Moreover, within the oligarch's group, transfer pricing is used to optimize profits and minimize taxes. \citet{gorodnichenko2008oligarchs} point out that the oligarchic holdings are better integrated with the world economy and are open to export of their production output.
In the model, we assume that the post-Soviet oligarch is a powerful entity having the following impact on the economic system:
\begin{itemize}
    \itemsep0.2em
    \item completely controls a set of companies $\mathcal{O} \subset M$; the set of companies not owned by the oligarch will be denoted by $\overbar{\mathcal{O}} = M \setminus \mathcal{O},$
    \item the set of companies $\mathcal{O}$ owned by an oligarch is a consistent part of the production chain (there are direct links between companies in $\mathcal{O}$ or they share the same raw materials),
    \item the goal of the oligarch is to maximize the value added of the companies in $\mathcal{O}$, that is $\sum_{m \in \mathcal{O}}{\rho_m} \rightarrow \max$ ,
    \item The oligarch has a significant influence on the political system and can influence the market regulator not to control oligarch's company $\mathcal{O}$ while simultaneously  providing the same regulatory incentives to not controlled by oligarch companies $\overbar{\mathcal{O}}$ that would have been taken under the optimal production plan $\psi^*$,
    \item The oligarch has a significant influence on the market and production chain and can limit transaction flows $\bm{x}$ between other market actors $\overbar{\mathcal{O}}$ when it is beneficial for the oligarch.
\end{itemize}

In this setting the goal function of the oligarch is defined as:
\begin{equation}
\psi_{\mathcal{O}} = \sum_{m \in \mathcal{O}}{
  \left(v_m \alpha_m \prod_{k \in K}{x_{km}^{\beta_{km}}}\right)} -
  \sum_{k \in K}{\left(v_k \sum_{m \in \mathcal{O}}{x_{km}}\right)}
\rightarrow \max
\label{eq:goalolig}
\end{equation}

Equation (\ref{eq:goalolig}) defines the value added of the set of companies $\mathcal{O}$ belonging to the oligarch as the difference between the value of their production output (the first summation) minus the required raw materials and intermediate products (the second summation).

The oligarch controls which transactions can be performed within the network. 
We measure the influence of the oligarch on the production chain by the parameter $\gamma \in [0,1]$ which defines the maximum percentage of a good that can be captured by the oligarch from the production chain -- further in the text we denote it simply as \emph{capture power}. That is, the oligarch captures goods that would otherwise be produced for the other companies to maximize its internal profit, possibly negatively affecting the total added value of the entire economy. $\gamma = 0$ means no capture of goods by the oligarch -- the oligarch considers in production planning only the goods that they would have available in the theoretical maximum value added $\psi^*$. On the other hand, $\gamma = 1$ means that the oligarch captures all goods produced by the non-oligarchic companies $\overbar{\mathcal{O}}$ and can use them for its own production. However, please note that, in both cases, regardless of the capture power $\gamma$, the oligarch is not interested in maximizing the overall value added to the economy, but only its own profit. In the real world, the strength of oligarch influence depends on their ability to shape policies \citep{winters2011oligarchy}, restrict access to critical resources and suppress new market entrants \citep{acemoglu2008oligarchic}, or use their economic influence \citep{hartmann2021hidden}.

Now, we can define the constraints for the oligarch's profit maximization defined in Equation (\ref{eq:goalolig}) as follows:
\begin{equation}
   \sum_{m_2 \in \mathcal{O}}{x_{m_1m_2}} \le \gamma \cdot y^*_{m_1} + (1-\gamma)\sum_{m_2 \in \mathcal{O}}{x^{*}_{m_1 m_2}} \qquad  \forall m_1 \in \overbar{\mathcal{O}}
\label{eq:balanceolig}
\end{equation}
This caps the resources available to the oligarch level in GDP-maximizing economy depending on the oligarch's capture power $\gamma$. 

We maximize the expression defined by Equations (\ref{eq:goalolig}) with regard to the oligarch's capture power for intermediate products in the economy, defined in Equation (\ref{eq:balanceolig}). We denote the solution of this maximization problem by $\psi_{\mathcal{O}}^*$. This is the maximum profit that the oligarch can obtain $\mathcal{O}$ within the economy represented by the graph $G$.

Once the oligarch with the capture power of $\gamma$ can achieve their optimal profit $\psi_{\mathcal{O}}^*$, we assume that the rest of the economy adapts to this situation. We define the second stage optimization problem: calculating the total added value (GDP) for the oligarch-controlled economy as:

\begin{equation}
    \psi_{\mathcal{O}}^* + \sum_{m \in \, \overbar{\mathcal{O}}}{
      \left(v_m \alpha_m \prod_{k \in K}{x_{km}^{\beta_{km}}}\right)} -
      \sum_{k \in K}{\left(v_k \sum_{m \in \, \overbar{\mathcal{O}}}{x_{km}}\right)}
    \rightarrow \max
    \label{eq:goalotherswtholig}
\end{equation}
subject to:

\begin{align}
\sum_{m \in \mathcal{O}}{
    \left(v_m \alpha_m \prod_{k \in K}{x_{km}^{\beta_{km}}}\right)} -
    \sum_{k \in K}{\left(v_k \sum_{m \in \mathcal{O}}{x_{km}}\right)} &\ge \psi_{\mathcal{O}}^*
\label{eq:minolig} \\
\sum_{m_2 \in M}{x_{m_1m_2}} - \alpha_{m_1} \prod_{k \in K}{x_{k{m_1}}^{\beta_{k{m_1}}}} &\le 0 \qquad \forall m_1 \in M
\label{eq:olgbalance}
\end{align}

Once the oligarch has ensured the flow of intermediate products and the production levels of $\mathcal{O}$ companies, the rest of the economy $\overbar{\mathcal{O}}$ adapts to this situation and maximizes the value added under the new conditions. The objective function in Equation (\ref{eq:goalotherswtholig}) assumes that the profit of the oligarch is already fixed at $\psi_{\mathcal{O}}^*$ and the rest of the economy can operate as long as the oligarch gets their $\psi_{\mathcal{O}}^*$ (see Equation \ref{eq:minolig}). Finally, Equation (\ref{eq:olgbalance}) ensures that the production plan is feasible with regard to the availability of semi-products.
 
\begin{figure}[ht!]
        \centering
        \includegraphics[width=0.7\textwidth]{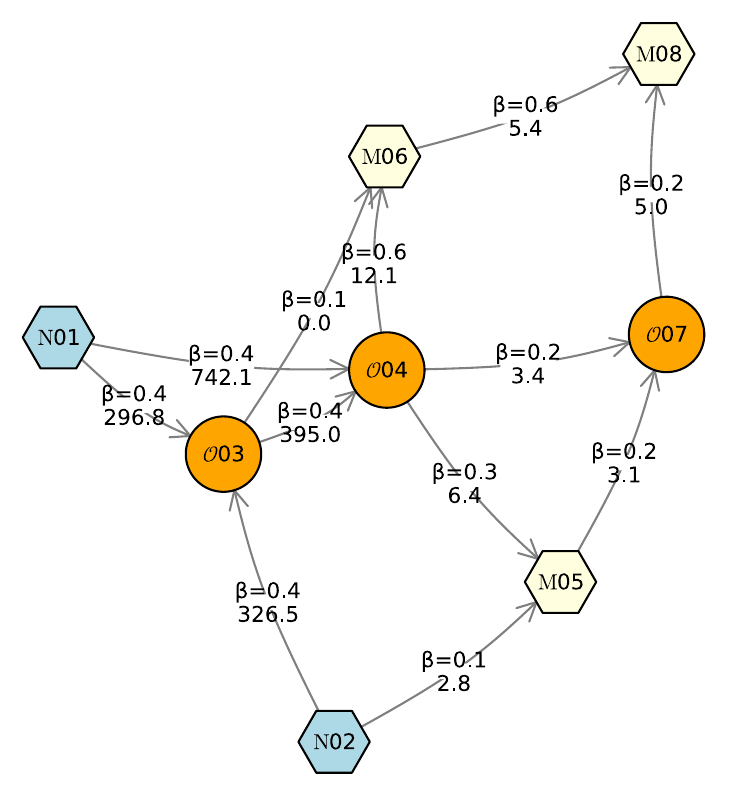}
\caption{The optimal production plan after the oligarch limited the transfer of goods within the network to maximize their profit from controlled nodes $\mathcal{O}03$, $\mathcal{O}04$, $\mathcal{O}07$. The nodes in the graphs controlled by the oligarch are orange circles.}
    \label{fig:production_chain_optimal_olig}
\end{figure}

Figure \ref{fig:production_chain_optimal_olig} shows a sample optimal production plan after the oligarch having the capture power of $\gamma = 1$ limited the transfer of goods within the network to maximize his profit from nodes $3$, $4$, $7$. It should be noted that in this numerical example under the original production plan the GDP was equal $\psi^* = 704.65$, while the profit of the oligarch was $\psi_{\mathcal{O}}^* = 640.83$. After the oligarch limited the transfer of goods within the network, their profit increased $643.61$, that is, by $2.77$ monetary units; however, the overall GDP of the economy decreased to $683.83$, that is, it fell by $20.82$ monetary units. This simple numerical example shows how inefficient the impact of the oligarch on the economy can be. We will investigate this further when discussing the results of numerical experiments in Subsection \ref{sec:resultgreedy}.

Let us now define the depth of the oligarch $d(\mathcal{O})$ in the production chain $G(K, E)$. Let us denote by $s(k_1, k_2)$ the length (measured by the number of edges) of the shortest path between the nodes $k_1$ and $k_2$ on the directed graph $G$. If no such path exists, we assume $s(k_1, k_2) = \infty$. The depth of the oligarch in the production chain is defined as the shortest path length between any natural resource node and any oligarch's node. That is:
\begin{equation}
    d(\mathcal{O}) = \min_{k_1 \in N, k_2 \in \mathcal{O}}{s(k_1, k_2)}.
\end{equation}

Please note that the oligarchs located at the beginning of the production chain are likely to be responsible for providing intermediate products to the companies later in the production chain. Hence, the trading decisions of these oligarchs are expected to have a significant impact on the overall economy. However, oligarchs located at the end of the production chain are responsible for providing final products for the market. Hence, their decisions might not affect other companies in the production chain. We will investigate this further through numerical experiments in Section \ref{sec:experiments}.

In the next section, we will conduct a series of numerical experiments that will allow us to understand how the size of the oligarch -- the number of oligarch companies $\mathcal{O}$, as well as the placement of the oligarch in the network, can influence the overall GDP of the economy.

\section{Experiments}
\label{sec:experiments}

The goal of this section is to present a series of Monte Carlo stochastic randomized numerical experiments on the proposed model of the oligarchic economy. We start with a description of the model implementation details and subsequently move to the experimental setup. Next, we present the results of the experiments and discuss them.

\subsection{Technical implementation details}
\label{sec:implementation}

We have implemented the model presented in Section \ref{sec:model} using the Julia 1.11.5 programming language developed by \cite{bezanson2012julia}. We use the Julia library JuMP.jl created by \cite{dunning2017jump}.
Please note that all models presented in Section \ref{sec:model} (that is, Equations (\ref{eq:goal}), (\ref{eq:goalolig}), (\ref{eq:goalolig})) have concave objective functions that are being maximized. Moreover, all constraints are convex or linear. Hence, this means that the feasible region of the optimization problem is convex. In turn, this means that the local optimum for the problems defined in Equations (\ref{eq:goal}), (\ref{eq:goalolig}), (\ref{eq:goalolig}) is also the global optimum. Hence, the optimal solution can be found using a local optimization solver such as Interior Point OPTimizer -- IPOPT \citep{wachter2006implementation}.

Another technical issue worth noting is related to numerical stability. The IPOPT solver requires that the optimized function is differentiable across the entire value domain. In our case, the production function defined in Equation~(\ref{eq:production_function}) is differentiable for all values of $x_{km} >0$. When defining the model for Ipopt.jl we have omitted in the objective function the terms with $\beta_{km} = 0$ as they have corresponding $x_{km} = 0$. For $\beta_{km}>0$, we have assumed $x_{km} \ge \epsilon$, where $\epsilon$ is a small value close to $0$. In this way, the differentiability was ensured, and IPOPT was able to find the optimal solution. Finally, for the model to be numerically solvable in the Equation~(\ref{eq:minolig}) the right-hand side of the inequality was set to $\psi_{\mathcal{O}}^* - \epsilon$. This setting allowed us to perform the numerical experiments discussed in the next subsection.

\subsection{Experiment setup}
\label{sec:experiment_setup}

The numerical experiments will be performed using Monte Carlo simulations over a population of $L = 10000$ synthetically generated economies. For the numerical experiments, we assume a fixed size of an economy with $|N| = 2$ natural resources and $|M| = 25$ companies. We assume that the vertex indices $K$ are topologically sorted. The values $\beta_{km}$ have been generated randomly uniformly from range $[0.25, 0.6]$ (with a step of $0.05$) in a way that satisfies the following conditions: 
\begin{itemize}
    \itemsep0.2em
    \item the number of incoming edges to companies $m \in M$ is $|\{k: \beta_{km} > 0\}| = 2$; 
    \item the economies of scale for for each company $m \in M$ are within the limit $\sum_{k \in K}{\beta_{km}} \in [0.5, 0.85]$;
    \item the graph $G$ crated from adjacency matrix $\bm{\beta}$ is a directed acyclic graph (DAG) and the depth of the graph $d(G)$ is at least $5$; 
    \item it is possible to generate an oligarch $\mathcal{O}$ having the size $|\mathcal{O}|$ at least $12$ nodes with the oligarch's depth $d(\mathcal{O})$ at least $3$. 
\end{itemize}
Setting the number of inputs to $2$ can be regarded as a simplified representation of oligarch-controlled and non-oligarch-controlled production inputs of a company. The assumptions about the economies of scale are typical for real-world markets. Representing the economy as a network of depth $5$ and accommodating oligarchs of size $12$ is a technical assumption that makes it possible to test various oligarchic structures within a single value-added network and achieve comparable results across different parametrizations.

For each company $m \in M$, the value $\alpha_m$ is randomly generated uniformly from the range $[1.1, 1.6]$ with a step of $0.1$ -- this represents heterogeneity of the technical level of companies within the network. Finally, we assume that the prices of goods are $v_k = 1.1^k$, $\forall k \in K$ -- since the vertex indices $k  \in K$ are topologically sorted, the value of goods increases with their depth in the production chain.
We will explore the impact of oligarchization through a set of $L = 1000$ randomly generated adjacency matrices $\beta_{km}$ representing production chains.

For each value-added network, represented by the graph structure $G$, we consider several possible oligarchic structures of different depths and sizes. We consider oligarch depths $1$--$5$ and for each such depth $d(\mathcal{O})$ we try to generate oligarchs of size $1$ up to $|M| - d(\mathcal{O})$. Depending on the network structure and the desired size of an oligarch, it will not always be possible to generate an oligarch with the desired properties.  As was mentioned earlier, an oligarch's sub-network is denoted by $\mathcal{O}$, its size $|\mathcal{O}|$ and its depth within the network $G$ by $d(\mathcal{O})$.
Moreover, we make sure that the nodes belonging to the oligarch are generated in a consistent way -- that is, they constitute a sub-graph which is also a DAG.

The detailed parameters of the simulation scenario have been collected in Table \ref{tab:sim_scenario}.

\begin{table}[ht!]
	\begin{centering}
		\begin{tabular}{| r | c | c |}
			\hline
			Parameter \qquad & Symbol   &   Values \\
			\hline\hline
			Number of natural resources \quad & $|N|$ & $2$ \\
            Total number of goods/companies \quad & $|M|$ & $25$ \\
            Size of the oligarch \quad & $|\mathcal{O}|$ & $1$--$25$ \\
            Oligarch's depth in the production chain \quad & $d(\mathcal{O})$ & $1$--$5$ \\
            Number of intermediate products & $|\{k: \beta_{km} > 0\}|$ & $2$ \\
            Economies of scale for companies $m \in M$ \quad & $\sum_{k \in K}{\beta_{km}}$ & $0.5$--$0.85$ \\
            Technology parameter \quad & $\alpha_m$ & $1.1$--$1.6$ \\
            Value of goods \quad & $v_k$ & $1.1^k$ \\
            Number of artificial economies generated \quad & $L$ & $1000$ \\
            Oligarch's capture power \quad & $\gamma$ & $0, \nicefrac{1}{4},\dots,1$ \\
			\hline
		\end{tabular}
	\end{centering}
	\caption{Parameters of the simulation scenario.}
	\label{tab:sim_scenario}
\end{table}

\subsection{Results}
\label{sec:results}

Following the experiment assumptions outlined in the previous subsection, we have run a battery of numerical experiments to show how the economic performance is affected by the oligarchization level.
We also show how the influence of oligarch on the economy depends on the depth $d(\mathcal{O})$ of their location in the production chain.
 
We will start by showing the damaging effect of the oligarch on GDP, next we move to the discussion of the oligarchic inefficiency and, finally, we will discuss the recommendations for the de-oligarchization process.
 
\subsubsection{Impact of oligarchs on GDP}
\label{sec:resultsdp}

We start the analysis of the results by showing the impact of oligarchs on the GDP of the economy. Throughout this subsection we assume that the oligarch has a full capture power -- that is $\gamma = 1$. The impact of various oligarch's capture power will be evaluated in the next subsection.

\begin{figure}
    \centering
    \includegraphics[width=0.7\textwidth]{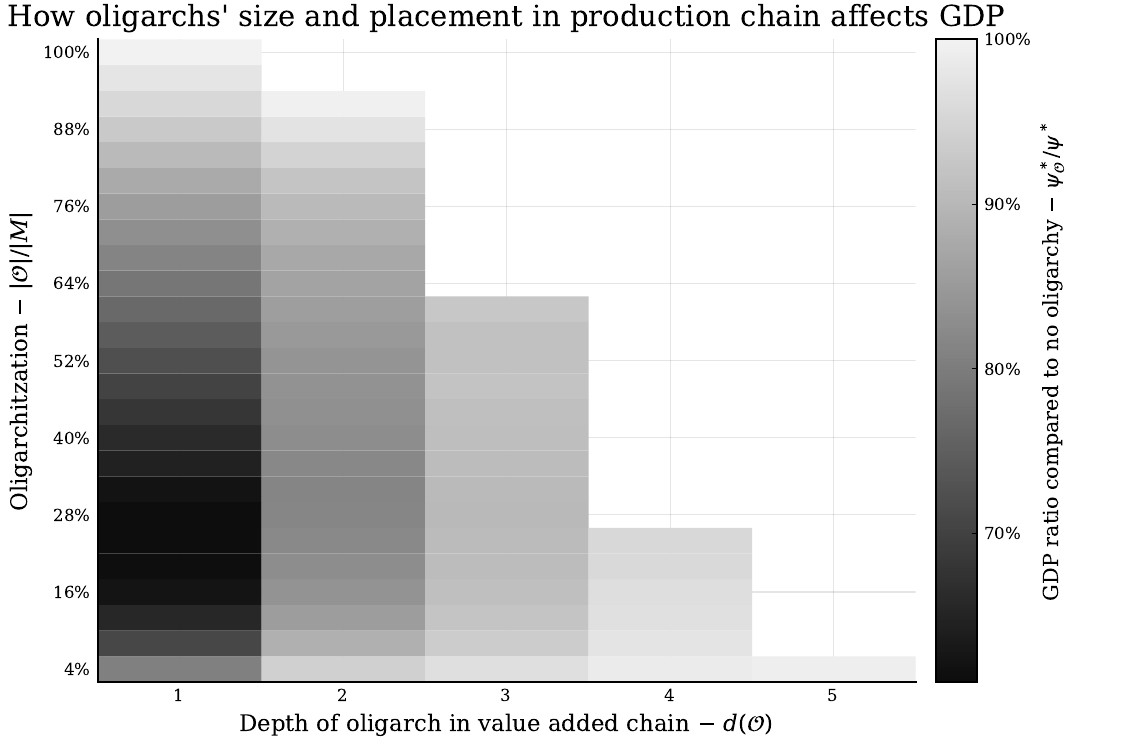}
    \caption{The impact of oligarchization on the GDP of the economy. The x-axis shows the depth of the  oligarch $\mathcal{O}$ in the value added chain and the y-axis shows the oligarchization level calculated as $|\mathcal{O}|/|M|$. The color denotes the GDP level compared to the optimal GDP $\psi^*$ without an oligarch. It can be seen that the crucial factor in de-oligarchization process is to increase the depth of the oligarch in the production chain as the oligarchs directly mining raw resources havethe most significant negative impact on the economy.}
    \label{fig:oligarchization_vs_gdp}
\end{figure}

The heatmap presented in Figure \ref{fig:oligarchization_vs_gdp} shows how the share of an oligarch in the economy, measured by the percantage controlled companies, $|\mathcal{O}|/|M|$ (y-axis) affects the overall GDP of the economy, depending on the depth of oligarchic structure $d(\mathcal{O})$ within the production chain (x-axis).

The data have been standardized to the \emph{``no oligarch''} case, that is the GDP of the economy without an oligarch is set to $100\%$ and the GDP of the economy with an oligarch is shown as a share of this value.  The relative GDP values are calculated as $\psi_{\mathcal{O}}^*/\psi^*$, where $\psi_{\mathcal{O}}^*$ is the GDP of the economy with an oligarch and $\psi^*$ is the optimal GDP without an oligarch. The color of the box represents the average relative GDP value $\psi_{\mathcal{O}}^*/\psi^*$  after running $L=1000$ stochastic simulation for various locations of oligarch in production chain $d(\mathcal{O})$ and various relative sizes of the oligarch $|\mathcal{O}|/|M|$. 
Note however, that, for some network structures, it was not possible to generate an oligarch with a sufficient depth $d(\mathcal{O})$ -- in such cases the average has been calculated for a smaller number of simulation runs.

It can be clearly seen that the severity of the impact of an oligarch over the economy strongly depends on the distance between the oligarch and natural resources. The oligarch of depth $d(\mathcal{O})=1$ (that is, the oligarch directly processing natural resources) of even moderate size has a very damaging impact on the output of economy. In The GDP of such an economy $\psi_{\mathcal{O}}^*$ can be below of $70\%$ of the optimal GDP $\psi^*$ of an economy without an oligarch (over 30\% GDP drop). This is because the oligarch is able to control all raw resources along with immediate intermediate products, thus limiting their availability to other companies in the production chain. The negative impact decreases with the depth of the oligarch in the production chain. The oligarch of depth $d(\mathcal{O})=2$ with the $30\%$ oligarchization decreases the GDP by about $15\%$ and the GDP decrease drops below $10\%$ for the depth $d(\mathcal{O})=3$. This leads to the conclusion that the de-oligarchization policies should have highest priority on the enterprises that are directly processing natural resources. 

Note, as pointed out previously, that in our model when an economy is fully controlled by an oligarch, its performance is equal to the theoretical maximum value added $\psi^*$. This is because in our approach the oligarch enforces the optimal production plan for the companies under their control. We do not include the ``oligarchical inefficiency'' factor. \citet{lankina2019soviet}, points out that the level of oligarchization in Ukrainian oblasts is positively correlated (0.777) with income per capita. However, later data from \citet{goriunov2023oligarchic} show a negative correlation between the level of oligarchization and GDP per capita in Ukrainian oblasts. In our model both cases are possible depending on the initial oligarchization level.

\subsubsection{Power capture of oligarchs}

In the previous subsection, we have assumed that the oligarch has a full capture power $\gamma = 1$. 
We are now going to drop this assumption and investigate how the oligarch's capture power $\gamma$ affects the overall GDP of the economy. We will consider the oligarchic capture values $\gamma = 0, \nicefrac{1}{4}, \nicefrac{1}{2}, \nicefrac{3}{4}, 1$. The value $\gamma = 0$ means that the oligarch cannot influence other transactions on the market (but still plans production only for theirs own profit) and the value $\gamma = 1$
means that the oligarch is capable to redirect all intermediate products to use for their own profit maximization -- see Equation \ref{eq:balanceolig}.

\begin{figure}
    \centering
    \includegraphics[width=0.7\textwidth]{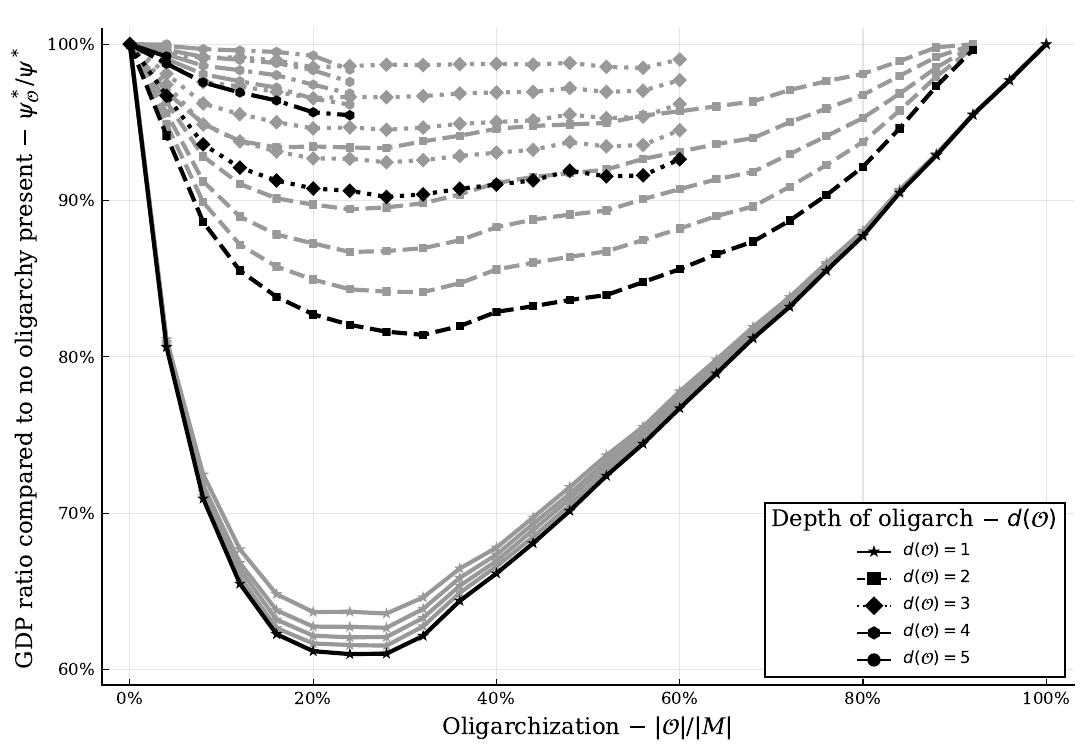}
    \caption{The impact of various capture power $\gamma$ and oligarchization on the GDP of the economy.  The x-axis shows the oligarchizaion level calculated as $|\mathcal{O}|/|M|$. The y-axis denotes the GDP level compared to the optimal GDP $\psi^*$ without an oligarch. 
    It can be seen that the crucial factor in de-oligarchization process is to increase the depth of the oligarch in the production chain.}
    \label{fig:oligarchization_impact_lines}
\end{figure}

The results presented in Figure \ref{fig:oligarchization_impact_lines} present various GDP drop levels depending on the size and and location of the oligarch in the production chain. The baseline scenario for full capture power $\gamma = 1$ is denoted with black lines -- their location corresponds to color intensity in the heatmap presented in Figure \ref{fig:oligarchization_vs_gdp}. The gray lines correspond to the lower levels of oligarchic capture $\gamma$. The upper gray line represent $\gamma = 0$, while the three middle lines represent intermediate capture levels $\gamma = \nicefrac{1}{4}, \nicefrac{1}{2}, \nicefrac{3}{4}$. 

It can be observed that for the oligarchs with the depth $d(\mathcal{O})=1$, that is oligarch having direct access to natural resources, the level of oligarchic capture $\gamma$ affects the production output only in a small degree (small spread of the bottom line). This happens due to the fact that the natural resource $n \in N$ can be acquired without limitations for a fixed price $v_n$. Since the raw resources constitute the main production input of the oligarch, the ability of an oligarch to maximize its profit is not heavily dependent on transactions taking place between non-oligarchic parts of the value-added chain.

On the other hand, when we consider an oligarch with depth $d(\mathcal{O}) \ge 2$, their capture power determines the GDP drop to a much greater degree. For an example, consider an oligarch with depth $d(\mathcal{O}) = 2$ and oligarchization level $|\mathcal{O}|/|M| = 30\%$. The oligarch having the full capture power $\gamma = 1$ (dashed black line with squares) will lead to a GDP drop of about $18\%$, while the oligarch of the same size without the capture power that is $\gamma = 0$ will lead to a GDP drop of only around $6\%$ (top gray dashed line for with squares). This is because the production output of the considered oligarch $d(\mathcal{O}) = 2$ is dependent on intermediate products that are also used by other companies in the production chain. Hence, the oligarch's ability to capture these intermediate products may limit their availability for other market actors which leads to a significant negative impact on the overall GDP of the economy. Similarly for the oligarch with depth $d(\mathcal{O}) = 3$ (dotted diamond line) with the oligarchization level $|\mathcal{O}|/|M| = 30\%$, the GDP drop is around $10\%$ for $\gamma = 1$ and only around $2\%$ for $\gamma = 0$.

The Ukrainian politicians as well as the international community have been aware of the negative impact of oligarchs on the economy for a long time \citep{gorodnichenko2008oligarchs}, \citep{pleines2016oligarchs}. Authors point out that the de-oligarchization is required to improve the economic performance of the country and accelerate its economic development 
\citet{rojansky2022} \citet{minakov2023war}. However, the results presented in Figure \ref{fig:oligarchization_impact_lines} show that the de-oligarchization process should be carefully planned. Note the convex shape of the lines for various oligarch depths and capture levels. The convex shape occurs due to the fact that when the oligarch controls almost entire production chain then their decisions are coherent with maximization of the output of the economy. However, the convex shape means that in in an economy with a significantly high level of oligarchization, the de-oligarchization process can lead to a GDP drop in the short term, especially for oligarchs with a direct access to raw resources (with depth $d(\mathcal{O})=1$). It should be noted though that this negative effect can be mitigated by cutting off oligarchs' access to natural resources as this means moving on a higher GDP curve in Figure \ref{fig:oligarchization_impact_lines}. In case of oligarchs not having direct access to natural resources (with depth $d(\mathcal{O}) \ge 2$) it is enough to limit their capture power $\gamma$ to achieve a significant increase in GDP.

In conclusion, a de-oligarchization strategy should firstly focus on changing the ownership of companies that are directly processing natural resources. The ownership change can happen e.g. via enforcing shareownership changes, divestiture, nationalization or regulatory changes (anti-trust laws, taxation, subsidizing new market entrants). In the second step there are two options: either to limit the oligarch's capture power $\gamma$ via enforcing better free-market mechanisms or further decreasing the share of oligarch in the economy, preferably in a way that leads to increase of their the depth in the production chain $d(\mathcal{O})$.

\subsubsection{Oligarchic inefficiency}
\label{sec:resultgreedy}

The results from the previous subsection show that de-oligarchization can lead to a significant increase in the GDP of the economy. The de-oligarchization may be performed by legal measures (strengthening rule of law, preventing corruption, asset seizure), changing the market rules (introducing anti-trust laws, taxation, subsidizing new market entrants) or by directly changing the ownership structure of the companies in the economy (e.g. via divestiture, nationalization, shareownership changes). Moreover, several reports point out that oligarchs take actions to block reforms related to privatization, energy trade, state procurement, and state aid -- for a discussion of Ukrainian oligarchs blocking market deregulations see  \citep{pleines2016oligarchs, pleines2008manipulating, stewart2013public, dimitrova2013shaping}. In conclusion, the de-oligarchization actions might require various types of compensations paid out to oligarchs. Hence this is very important to understand the cost-benefit ratio of de-oligarchization.

In this section, we will investigate how much of the money that the oligarch is taking from the economy is actually going to the oligarch's pocket and how much is lost in the inefficiency of the economic allocation.

\begin{figure}
    \centering
    \includegraphics[width=0.7\textwidth]{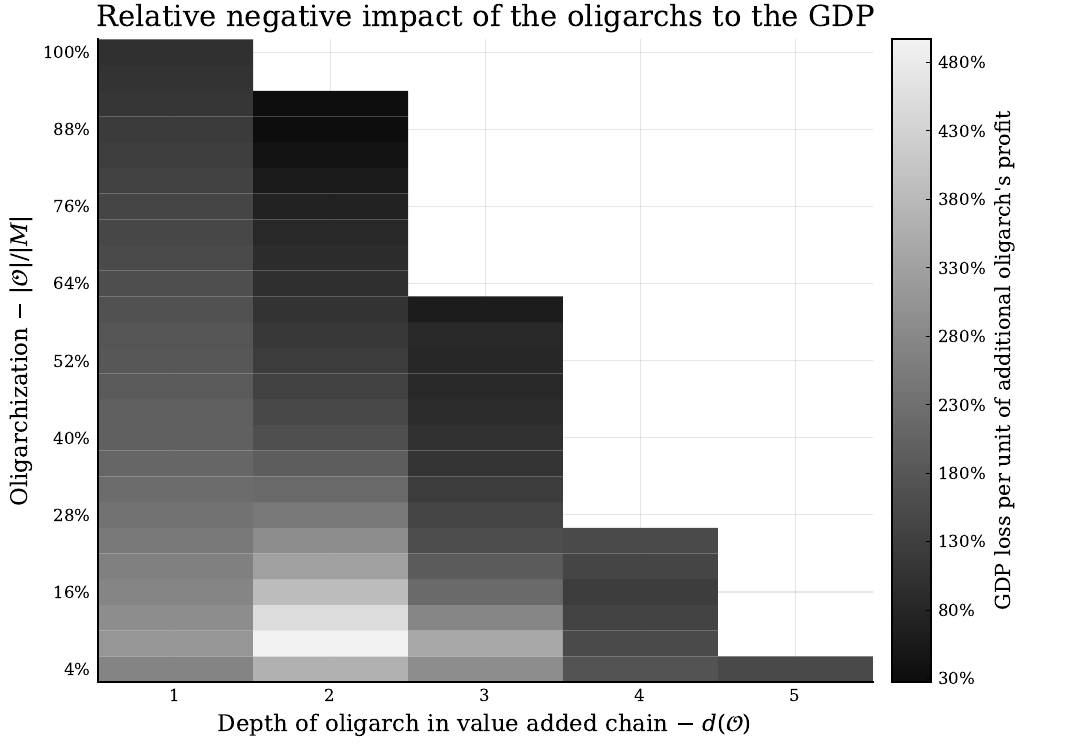}
    \caption{The de-oligarchization could lead to significant relative economic gains that depend on the location of the oligarch in the production chain. The x-axis shows the depth of the  oligarch $\mathcal{O}$ in the value added chain and the y-axis shows the oligarchizaion level calculated as $|\mathcal{O}|/|M|$. The color denotes how many units of GDP were scarified in order to increase the profit of the oligarch by 1 unit.}
    \label{fig:oligarch_greedeness_vs_gdp}
\end{figure}

Figure \ref{fig:oligarch_greedeness_vs_gdp} shows the relative negative impact of the oligarchs to the GDP of the economy.
Each cell in the heatmap represents the number of units of GDP that were sacrificed in order to increase the profit of the oligarch by 1 unit. As we have shown in the previous subsection, the oligarch, via his actions, is decreasing the overall efficiency of the economy via inefficient allocation of resources and intermediate products within the production chain. This means that increase of oligarch's profit (compared to the situation where there was no oligarch) always leads to a decrease of GDP. What is stunning is the scale of that decrease, which can be up to $4.8\times$. This means that in some cases increasing the profit of oligarch by \$1  leads to the decrease of GDP by $\$4.8$.
Note that the total damage by the oligarch to the economy is illustrated in Figure \ref{fig:oligarchization_vs_gdp}. An oligarch of depth $d(\mathcal{O})=2$ and size $|\mathcal{O}|=2$ will reduce GDP by roughly $10\%$. However, as shown in Figure \ref{fig:oligarch_greedeness_vs_gdp}, only a small fraction (approximately $1/4.8 = 21\%$ of this money goes to the oligarch's pocket. The rest is lost in the inefficiency of new economic allocation as a result of oligarch's actions.
If the regulator is collecting some part of companie's profit as a tax, then the regulator could even afford to pay the oligarch \$1 with taxes collected from the \$3.8 gain in the GDP value. 

An important observation is that why depth oligarchs $|\mathcal{O}|=2$ are the most inefficient to the economy in terms of the damage they do to GDP versus the money they recover as their profit. This is because their impact is two-fold. Firstly, they force companies at depth $1$ (that is, companies that have direct access to raw resources) to redirect their production for the benefit of the oligarch. This greatly reduces the availability of intermediate products for the rest of the economy. Secondly, they are still deep enough in the value-added chain to be able to heavily affect the performance of the overall economy by their actions. 

In conclusion, for some production chains, even if the oligarch is to receive a full compensation for their lost profit, the regulator could finance it with just the taxes collected from the increased GDP. This means that the de-oligarchization process could be self-financing by gains in efficiency of the overall economy.

\subsection{Discussion}
\label{sec:discussion}

The presented results show the negative impact of oligarchs on the economy depending on their size, location in production chain and capability to control other transactions on the market. However, the results obtained can be used to identify recommendations for the de-oligarchization process.

The results from the numerical experiments lead to the following recommendations for the de-oligarchization policy:
\begin{itemize}
    \item Prioritize the removal of oligarchs from companies that are directly processing raw materials, as their control has the most significant negative impact on the economy.
    \item If the oligarch is located in the middle or towards the end of the production chain, then it is enough to limit their capture power $\gamma$ to achieve a significant increase in GDP.
    \item Each unit of additional oligarch's profit can cost several units of lost GDP. Hence, in some scenarios (especially for small sized oligarchs) the de-oligarchization process can be self-financing by gains in efficiency of the overall economy. 
    \item If the initial level of oligarchization of economy is particularly high, then the de-oligarchization process can lead to a short-term decrease in GDP. This can be observed especially if the oligarch is located at the beginning of the production chain. However, this negative impact can be partially mitigated by starting the de-oligarchization process from the companies that are directly processing raw materials.
\end{itemize}

The results from the numerical experiments clearly show that the most efficient way to de-oligarchize the economy is to start from the companies that are directly processing raw materials $d(\mathcal{O})=1$. The control of companies close to the root of the production chain is the most damaging to the economy, as this affects all subsequent stages within the production chain. This conclusion is consistent with the litterature poinint out that oligarchs strengten their economic dominance by restricting access to critical resources such as financial capital, infrastructure, and technology -- \citep{acemoglu2008oligarchic}.

Once the natural resources are eliminated from the oligarchic chain, the next step is to limit the oligarchic capture power that influences the ability of other market actors to conclude transactions. Following \citet{robinson2012nations} oligarchs will prioritize rent-seeking over economic progress, using their influence to block competition and secure monopolistic privileges.  The oligarchs might also exploit the limited human resources availability as pointed out in \citep{matuszak2012democracy}.  From our results, the oligarch's capture power (represented by as as $\gamma$) is the second most damaging factor to the economy after the oligarch's control over natural resources. 
The initial actions leading to limiting the oligarchic power of controlling the economy might include the elimination of corruption and the strengthening of the rule of law \citep{kuznetsov2023ukraine}. Furthermore, The oligarchs' negative influence on market competition can be limited by introducing anti-trust laws, taxation, subsidizing new market entrants or by directly changing the ownership structure of the companies in the economy (e.g. via divestiture, nationalization, shareownership changes). 

Once the initial regulatory measures are taken and the oligarchization level is lower, the further de-oligarchization might require providing some compensations to the oligarchs. However, for lower oligarchization level the slope in Figure \ref{fig:oligarchization_impact_lines} is very sharp which in turn results in high economic gains of de-oligarchizations as discussed in \ref{fig:oligarch_greedeness_vs_gdp}. This means that in the final stage of de-oligarchization, the compensation for remaining oligarchs could be financed by additional tax yields due to more efficient de-oligarchized economy.

\section{Conclusions}
\label{sec:conclusions}

In this paper, we have presented a model of an economy with a production chain partially controlled by an oligarch. The model is based on the value-added network approach where outputs of companies are treated as intermediate products for producers further in the production chain. The model allows us to analyze the impact of oligarchs on the overall efficiency of the economy by introducing an oligarchic sub-network that can control a fraction of the market. Additionally, the model allows us to analyze the oligarch's ability to influence transactions between other market actors (which we call the oligarch's capture power). The proposed model has been implemented in the Julia programming language and has been tested in a series of randomized numerical experiments across various synthetic economies.

The results show that an oligarch having even a small fraction of the market under his control can lead to a critical decrease of economy efficiency. On the other hand, we have also shown that in economies with high levels of oligarchization, decreasing the role of oligarch can initially lead to the reduction of the overall GDP of the economy. This negative impact can be mitigated by starting de-oligarchization from the companies closer to the root of the production chain (e.g., mining raw resources).

Another important finding is that the ability of an oligarch to influence the transactions between other market actors (which we named the oligarch's capture power) by  affecting the efficiency of semi-product allocation can have a very significant damaging effect for the economy. Hence, alongside de-oligarchization of raw materials mininig, the second most important focus for policymakers is to limit oligarchs' ability to control market transactions. This can be achieved by reducing corruption, taxation rules for oligarchic chains, introducing anti-trust laws as well as  subsidizing new market entrants. 

It should also be noted that the negative impact on the economy is so significant that in some scenarios the potential cost of de-oligarchization can be self-financing just by the gains from higher taxes collected due to GDP growth. This means that the compensation costs of nationalization of oligarch-controlled assets could be covered by the economic gains. This statement is particularly true for the branches of industry with a smaller degree of oligarchization.

The presented model has several limitations. Firstly, we have assumed that the production technology across the chain is fixed and the $\beta_{km}$ parameters are constant. Secondly, in the proposed model the production chain is a directed acyclic graph, while in the real world there might be loops in the production chain. Thirdly, we have assumed decreasing economics of scale. Lastly, we have assumed that the prices of goods $v_k$ are static and are not influenced by the flow or possible importance of the goods within the chain. This situation could be found on a market where excessive production of possible because it could be exported on international markets. However, in further research, we are going to lift that assumption and introduce a dynamic pricing mechanism together with other economic of scale models.

Another limitation of the presented results is that the numerical experiments used synthetically generated economies. However, the proposed model could be calibrated to real-world economies using data on production chains and oligarchic structures. Hence, it can be applied to real-world economies, such as the Ukrainian economy, to identify the most efficient de-oligarchization policies. Even by running simulations on synthetically generated economies across randomized scenarios, we can demonstrate the mechanisms behind the negative impact of oligarchization on the economy and provide recommendations for de-oligarchization. Hence further research plan includes collecting real-world data
 to refine the model and use it for a more precise estimation of the outcomes of policy interventions.

\vspace{1em}
\noindent\textbf{Acknowledgments}\\
This research was supported by the National Science Centre, Poland grant number 2021/41/B/HS4/03349. For the purpose of the Open Access policy, the
authors applied a CC-BY 4.0 public copyright license to this preprint.

\vspace{1em}
\noindent\textbf{Declaration of generative AI and AI-assisted technologies in the writing process.}\\
During the preparation of this work the authors used the following LLM models: Overleaf Writefull and OpenAI's GPT in order to improve the readability and language of the manuscript, correct grammatical and spelling errors errors and enhance overall clarity. After using this tools, the authors reviewed and edited the content as needed and take full responsibility for the content of the published article.


\bibliographystyle{elsarticle-harv}

\bibliography{bibliography}

\end{document}